\documentclass[conference]{IEEEtran}
\IEEEoverridecommandlockouts
\IEEEpubid{\makebox[\columnwidth]{978-1-7281-8674-0/21/\$33.00~\copyright 2021 IEEE \hfill} \hspace{\columnsep}\makebox[\columnwidth]{ }}

\usepackage{cite}
\usepackage{amsmath,amssymb,amsfonts}
\usepackage{algorithmic}
\usepackage{graphicx}
\usepackage{textcomp}
\usepackage{xcolor}
\usepackage{orcidlink}
\usepackage{braket}
\usepackage{listings}
\usepackage{xcolor}
\usepackage{hyperref}
\usepackage{todonotes}
\usepackage{subcaption}
\usepackage{svg}

\definecolor{codegreen}{rgb}{0,0.6,0}
\definecolor{codegray}{rgb}{0.5,0.5,0.5}
\definecolor{codepurple}{rgb}{0.58,0,0.82}
\definecolor{backcolour}{rgb}{0.95,0.95,0.92}

\newcommand{\rosnet}{\texttt{RosneT}}

\lstdefinestyle{mystyle}{
    backgroundcolor=\color{backcolour},
    commentstyle=\color{codegreen},
    keywordstyle=\color{magenta},
    numberstyle=\tiny\color{codegray},
    stringstyle=\color{codepurple},
    basicstyle=\ttfamily\footnotesize,
    breakatwhitespace=false,
    breaklines=true,
    captionpos=b,
    keepspaces=true,
    numbers=left,
    numbersep=5pt,
    showspaces=false,
    showstringspaces=false,
    showtabs=false,
    tabsize=2
}
\def\BibTeX{{\rm B\kern-.05em{\sc i\kern-.025em b}\kern-.08em
    T\kern-.1667em\lower.7ex\hbox{E}\kern-.125emX}}

\graphicspath{{images/}}

\begin{document}

\title{\rosnet: A Block Tensor Algebra Library for Out-of-Core Quantum Computing Simulation}

\author{
\IEEEauthorblockN{
    Sergio Sanchez-Ramirez~\orcidlink{0000-0003-4273-3328}\IEEEauthorrefmark{1},
    Javier Conejero~\orcidlink{0000-0001-6401-6229}\IEEEauthorrefmark{2},
    Francesc Lordan~\orcidlink{0000-0002-9845-8890}\IEEEauthorrefmark{2},
    Anna Queralt~\orcidlink{0000-0003-2782-2955}\IEEEauthorrefmark{2}\IEEEauthorrefmark{3},
    Toni Cortes~\orcidlink{0000-0002-2537-8937}\IEEEauthorrefmark{2}\IEEEauthorrefmark{3},
}
\IEEEauthorblockN{
    Rosa M Badia~\orcidlink{0000-0003-2941-5499}\IEEEauthorrefmark{2},
    Artur Garcia-Saez~\orcidlink{0000-0003-3561-0223}\IEEEauthorrefmark{1}
}
\IEEEauthorblockA{\IEEEauthorrefmark{1}
\textit{QUANTIC, Barcelona Supercomputing Center}, Barcelona, Spain
\\\{sergio.sanchez.ramirez, artur.garcia\}@bsc.es}
\IEEEauthorblockA{\IEEEauthorrefmark{2}
\textit{Workflows and Distributed Computing, Barcelona Supercomputing Center}, Barcelona, Spain
\\\{javier.conejero, francesc.lordan, anna.queralt, toni.cortes, rosa.m.badia\}@bsc.es}
\IEEEauthorblockA{\IEEEauthorrefmark{3}
    \textit{Universitat Polit\`ecnica de Catalunya},
    Barcelona, Spain}
}

\maketitle
\IEEEpubidadjcol

\begin{abstract}
    With the advent of more powerful Quantum Computers, the need for larger Quantum Simulations has boosted. As the amount of resources grows exponentially with size of the target system Tensor Networks emerge as an optimal framework with which we represent Quantum States in tensor factorizations. As the extent of a tensor network increases, so does the size of intermediate tensors requiring HPC tools for their manipulation. Simulations of medium-sized circuits cannot fit on local memory, and solutions for distributed contraction of tensors are scarce. In this work we present \rosnet, a library for distributed, out-of-core block tensor algebra. We use the PyCOMPSs programming model to transform tensor operations into a collection of tasks handled by the COMPSs runtime, targeting executions in existing and upcoming Exascale supercomputers. We report results validating our approach showing good scalability in simulations of Quantum circuits of up to 53 qubits.
\end{abstract}


\section{Introduction} \label{section:introduction}
With recent Quantum Devices showing increasing capabilities to perform controlled Quantum operations, further development on Quantum Algorithms may benefit from Quantum Simulations on classical hardware. Among important applications one finds verification and debugging of Quantum Algorithms, and elucidating the frontier for real Quantum Advantage of new devices \cite{Villalonga_2020}. Due to the exponential cost of running exact simulations of Quantum systems, these simulations are designed for large computer clusters.

Among different approaches to the simulation of Quantum systems, Tensor Networks are regarded as an efficient numerical representation of a Quantum Circuit\cite{markov_simulating_2009}, allowing the simulation of large Quantum circuits. Numerical techniques based on Tensor Networks have been extensively applied to the study of Quantum systems but in the general case of simulating an arbitrary Quantum circuit however, the exponential growth forces tensors to be distributed among computing nodes. A number of methods and libraries have appeared recently to implement Quantum Simulators with Tensor Networks \cite{solomonik_cyclops_nodate,lyakh_domainspecific_2019,nguyen_tensor_2021,Wang_2021} intended for HPC clusters.

In this work we develop a Python library called \rosnet\ using a task-based programming model able to extend all tensor operations into distributed systems, and prepared for existing and upcoming Exascale supercomputers It is compatible with the Python ecosystem (\texttt{numpy}, \texttt{opt-einsum}, \texttt{quimb} and \texttt{cotengra}), and offers a simple programming interface for developers.

\rosnet\ is developed on top of PyCOMPSs~\cite{tejedor_pycompss_2017}, a task-based programming model for distributed computing. Between the features that favor the usage of this environment we find the possibility of specifying the resources needed for a given task (i.e., number of computing cores), an automatic out-of-core memory management that enables large-data applications, and the possibility of task polymorphism, enabling more than one version of the tasks specialized for different hardware (i.e., CPU versus GPU version).

While available options for simulation mostly store data in memory and do not exploit available disk space, storing data on disk has a pair of advantages:
\begin{enumerate}
    \item Clusters have more disk space than memory, so bigger quantum states can be represented. Memory is then fully dedicated to computation and caching.
    \item Non-volatile storage makes the simulation fault-tolerant with minimal redundancy. This is critical for cluster-wide executions, where thousands of nodes are used and crashes are frequent.
\end{enumerate}

\rosnet\ aims to bring all these features through a Python programming environment to a wide range of scientists developing Quantum simulations.

This work is organized as follows: Section \ref{section:methods} introduces the main operations used in this paper, namely Tensor Networks contractions. Section \ref{section:pycompss} describes the PyCOMPSs programming model. The implementation of our method is presented in Section \ref{section:implementation} and its evaluation is presented in Section \ref{section:evaluation}. Next, Section~\ref{section:related_work} provides a brief overview of the current state of the art. Finally, Section \ref{section:conclusions} wraps up the paper with the conclusions and future work.

\section{Methods} \label{section:methods}
\subsection{Tensors as Quantum States}

A Quantum State $|\Psi\rangle$ describes the state of a Quantum system along a Quantum computation, and its modifications by applying Quantum gates. $|\Psi\rangle$ is a state in a Hilbert space of dimension $2^n$, where $n$ is the number of qubits of the system. An explicit representation of the state is an  $n$-order unit tensor on $\bigotimes^n_{i=1} \mathbb{C}^2$ where  $\mathbf{i} = i_1 \dots i_n$ are the indices of the qubits. This tensor can be further decomposed in a structured composition of smaller tensors $A_k$ forming a Tensor Network. Using a general tensor contraction operation $\Phi$ we define the state as
$$|\Psi\rangle = \sum_{i_1 \dots i_n}\Phi(A_k) |i_1 \dots i_n\rangle.$$
Quantum Gates acting over Quantum States are also represented as tensors of the network, such that any physical observable and probability outcomes are obtained solely from operations over the tensors $A_k$. $1-$qubit gates are rank$-2$ matrices while $2-$qubit gates are expressed as tensors $C^{ij}_{kl}$, using 4 rank$-2$ indices.

While the creation and storage of the Tensor Network can be obtained in a straightforward manner from the Quantum circuit description, perform the contraction operation $\Phi(A_k)$ is a complex task in the general case, in fact obtaining the best contraction sequence is a hard problem \cite{g_a_smith_opt_einsum_2018,gray_hyper-optimized_2021}.

The contraction is designed as a sequence of operations between neighboring tensors creating new intermediate tensors, which may grow in size as the resulting tensor of a contraction operation has the combination of outward indices not contracted in the operation, \emph{i.e.}
$$
    A^{i_1\ldots i_k}_m B^m_{j_1\ldots j_r} = C^{i_1\ldots i_k}_{j_1\ldots j_r}.
$$
This imposes limits on the size of the tensor networks that we can contract, as the size of tensor grows exponentially fast with its order, and we expect large tensors to appear in intermediate steps of the overall contraction.

\subsection{Slicing}
A common method for handling large tensors is tensor cutting or slicing \cite{villalonga_flexible_2019,markov_quantum_2018,chen_classical_2018}.
Slicing is based on the idea of Feynman's Path Integral formulation of quantum mechanics, that all the paths in a tensor network equally contribute to the final quantum state.
An index (or edge) in the tensor network represents a vector space of dimensionality $D$ between two tensors. If we select one and project it into one of its basis, then we are filtering $\frac{1}{D}$ of the paths and the intermediate tensors that contain the index will shrink to occupy a $\frac{1}{D}$ of its original space at the cost of $\frac{1}{D}$ fidelity in the final result. Furthermore, if we make copies of the tensor network, each projected to a different basis of the selected index, and repeat this process, we end up with a massively parallel problem and can calculate up to the desired fidelity. However, contraction of non-sliced tensors and sliced tensors shared between instances become redundant and may add a significant overhead to computation. We notice that by viewing tensor projections as block divisions of the original tensors, this compute overhead disappears and becomes an indicator of block reuse while still can fit in a distributed system.

\subsection{Block Tensor Contraction}
Tensors used in Quantum Computing and Quantum Information fields usually have an index dimensionality of $2$ and the space occupied is of $2^n$ elements where $n$ is the order of the tensor. Thus a standard compute node with 100~GB of main memory may fit tensors of up to order 33 using single-precision complex numbers.
Larger tensors need to be sliced and distributed to different nodes.
Based on the Dimension Extents layout \cite{lyakh_domainspecific_2019,solomonik_cyclops_nodate}, we slice tensors into uniform grids of sub-tensors or blocks.

\begin{figure}
    \centering
    \includegraphics[width=.45\textwidth]{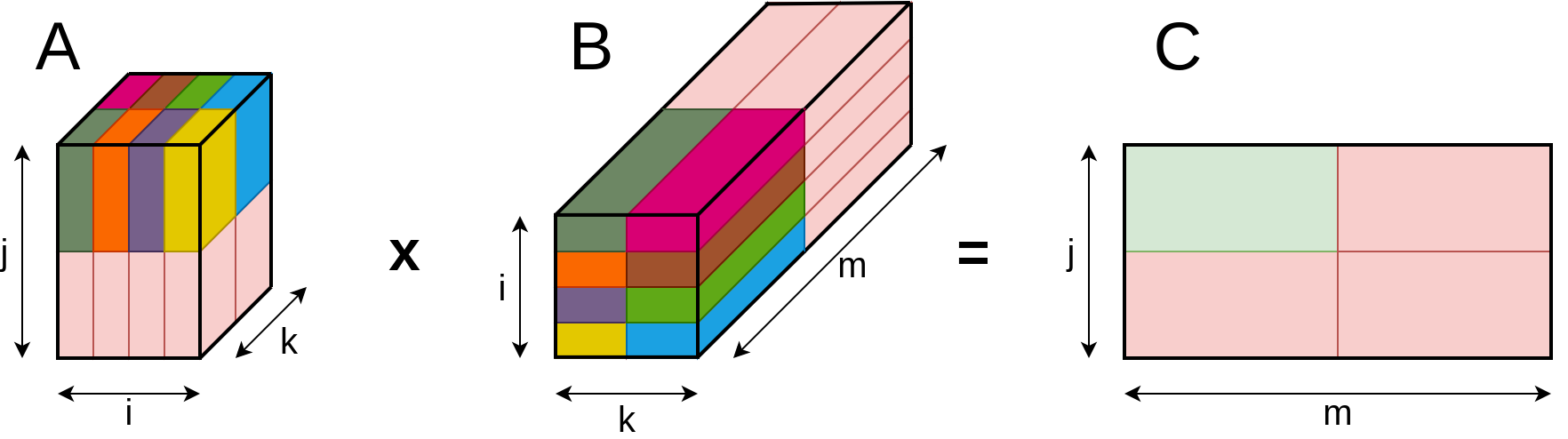}
    \caption{Example of a block tensor contraction between two 3-order tensors into a 2-order tensor. To compute a block of $C$ (light green), the needed pairwise $A,B$ blocks are colored in a range of colors by matching couples.}.
    \label{fig:pairwise-coupling}
\end{figure}
\begin{figure}
    \centering
    \includegraphics[width=.45\textwidth]{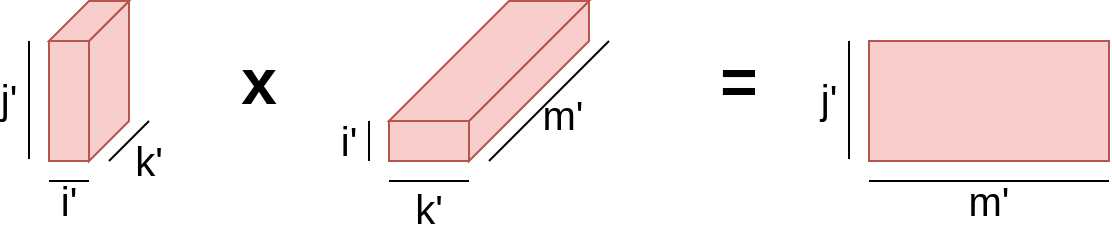}
    \caption{Tensor contraction of a pair of coupled blocks. The resulting tensor is a partial contribution of the final tensor.}
    \label{fig:pairwise-block-contraction}
\end{figure}
\begin{figure}
    \centering
    \includegraphics[width=.45\textwidth]{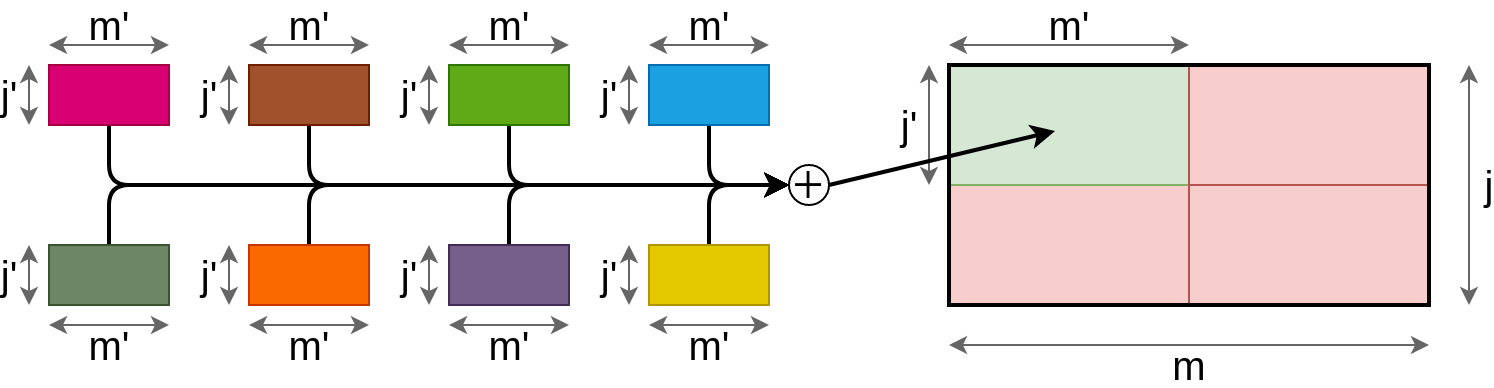}
    \caption{Reduction of partial block contractions into the final block tensor.}
    \label{fig:partial-sum}
\end{figure}

Employing the example in Figure~\ref{fig:pairwise-coupling}, block tensor contraction between tensors $A_{ijk}$ and $B_{ikm}$ is performed following these steps:
\begin{enumerate}
    \item For each output block in tensor $C$, select the blocks from $A$ and $B$ needed to compute the resulting block. These blocks are chosen by matching coordinates from non-contracting indexes $j,m$.
    \item Once the input blocks of $A$ and $B$ have been selected, group them in pairs by matching the coordinates of contracting indexes $i,k$ (Figure~\ref{fig:pairwise-coupling}).
    \item Each of the pairwise coupled blocks, when contracted locally, will compute a partial contribution to the output block (Figure~\ref{fig:pairwise-block-contraction}).
    \item Partial result blocks are finally added into the block of $C$ (Figure~\ref{fig:partial-sum}).
\end{enumerate}

\section{PyCOMPSs programming model} \label{section:pycompss}

PyCOMPSs is a task-based programming model that enables the parallel execution of sequential codes in distributed computing platforms, and in particular in large supercomputing clusters ~\cite{tejedor_pycompss_2017}. The parallelism is obtained from the parallel execution of tasks, that are methods identified by the programmer with the {\tt @task} decorator. For each task, the directionality of the method parameters is also indicated whether it is read or written. At execution time, this information is used by the runtime (COMPSs ~\cite{badia_comp_2015}) to detect the data dependencies between tasks and to build a Direct Acyclic Graph (DAG) where each node denotes a task instance and edges denote data dependencies between them. See listing \ref{lis:sample-task} for an example of annotated task.

\begin{lstlisting}[caption={Sample task annotation},captionpos=b,language=Python, label={lis:sample-task}]
@constraint(computing_units="2", memory_size="4")
@task(c=INOUT, a=IN, b=IN)
def madd(c, a, b):
  c += a * b
\end{lstlisting}

On top of the {\tt @task} decorator, PyCOMPSs supports other annotations that enable to specify multiple features. For example, the {\tt @constraint} decorator can be used to indicate hardware or software requirements of the task. In the sample task shown in listing
\ref{lis:sample-task} the task is requesting 2 cores ({\tt computing\_units}) and 4 GB of memory as resources needed to execute the task. This information is used by the COMPSs runtime to allocate the corresponding resources in the computing node.

\begin{figure}
    \centering
    \includegraphics[width=.45\textwidth]{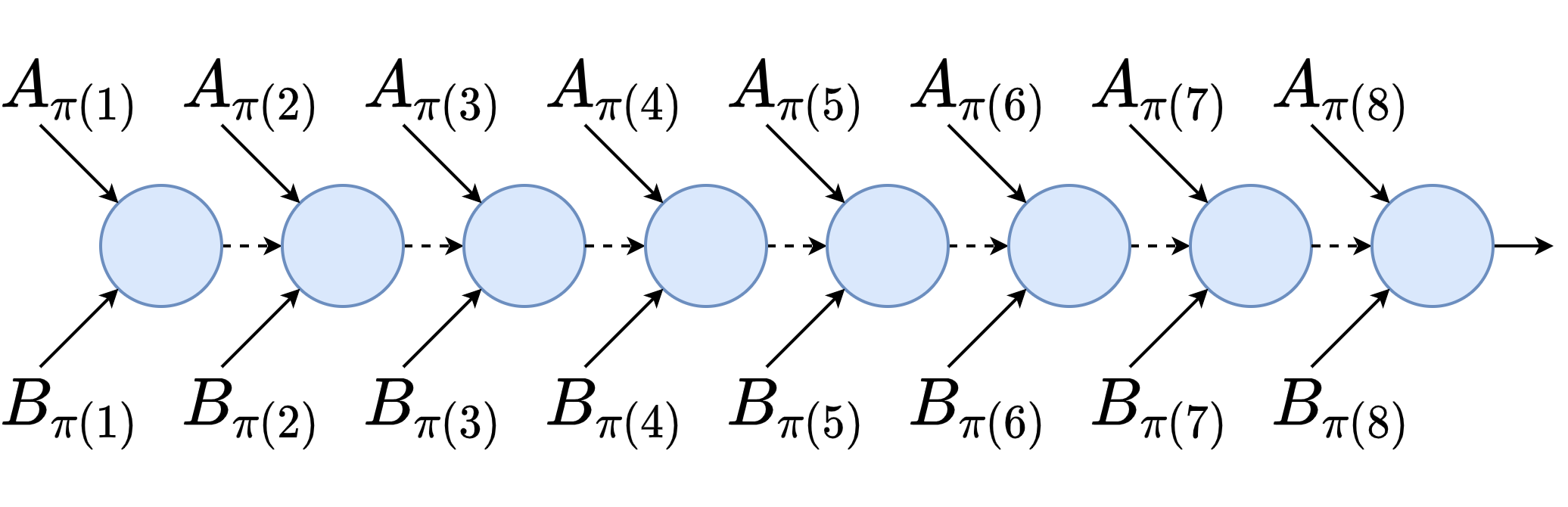}
    \caption{Task graph of the commutative reduction implementation. The ordering $\pi$ of the tasks is decided at runtime by the scheduler. Dashed arrows indicate the commutative dependency of the \textit{accumulator} variable.}
    \label{fig:commutative-reduction}
\end{figure}

While the relative order of tasks identified by the runtime in the application DAG is respected at execution time, there can be relaxations. One of the relaxations is done through the use of the {\tt COMMUTATIVE} clause (see listing \ref{lis:comm-task}) on the task parameters. Given a set of tasks that operate on a value tagged as commutative parameter, the runtime can opt to execute the tasks in whichever order it considers is better, but never executing more than one task at a time. Figure \ref{fig:commutative-reduction} shows a task graph for a set of tasks that perform a reduction operation on a {\tt   commutative} parameter that is read and written by each task. If implemented without the commutative option, these tasks would be executed one after the other in order of invocation. With the commutative clause, the runtime can opt to execute this set of tasks in any order, launching those whose input parameters are available earlier, enabling a faster execution of the task set.


\begin{lstlisting}[caption={Sample commutative task},captionpos=b,language=Python, label={lis:comm-task}]
@constraint(computing_units="24", memory_size="45")
@task(res=COMMUTATIVE)
def commutative_24(res, a, b, axes):
    res += np.tensordot(a, b, axes)
\end{lstlisting}

Another decorator included in the PyCOMPSs syntax is the {\tt implements} one, which enables to provide multiple versions of a task. These multiple versions can provide different implementations of the same behavior, and combined with the {\tt constraint} decorator can provide versions tailored to different hardware. For example, one can consider on offering a version of the task tailored for CPUs and another one for GPUs. The runtime will choose at execution time which version to use, in case the specific hardware (i.e., GPUs) is available.

The COMPSs runtime is deployed in the computing infrastructure following the master-worker paradigm, where the master node hosts the main program and the runtime and the worker nodes execute the tasks. The main program starts its execution in the master node and, each time a task is called, the runtime receives an invocation infers the dependencies of the current task, and the task is added to the task graph. Thus, the task graph represents a partial order of the application execution, where its potential concurrency is inherent. The runtime is responsible for deciding which tasks to execute at each moment, taking into account the existing data dependencies, task constraints and availability of the resources.

Another feature of PyCOMPSs is that it offers the illusion of a single shared memory and storage space.
Data is allocated in the node where it is first defined, but the runtime tracks its location and performs the corresponding data transfers between nodes when necessary. In addition, data parameters in PyCOMPSs only reside in memory of the node while the parameter is needed for a task execution. Upon the task completion, the runtime serializes the results into files. Similarly, before each task execution, the runtime transfers to the correspoding worker file system the necessary files and deserializes the input parameters. While this serialization/deserialization activity can be seen as an overhead, it allows
PyCOMPSs to implement an out-of-core execution model that enables the execution of very large data problems~\cite{alvarez_cid-fuentes_dislib_2019}. The only limitation is that the size of the task parameters fits in the memory of one node, and that the whole data set fits in the storage system. To reduce the number of file transfers between nodes, data locality is taken into account by the runtime, scheduling tasks in the nodes where data is when possible.

\section{Implementation} \label{section:implementation}
\subsection{Block Tensor Contraction with Commutative Reduction}

\begin{figure}
    \centering
    \includegraphics[width=.35\textwidth]{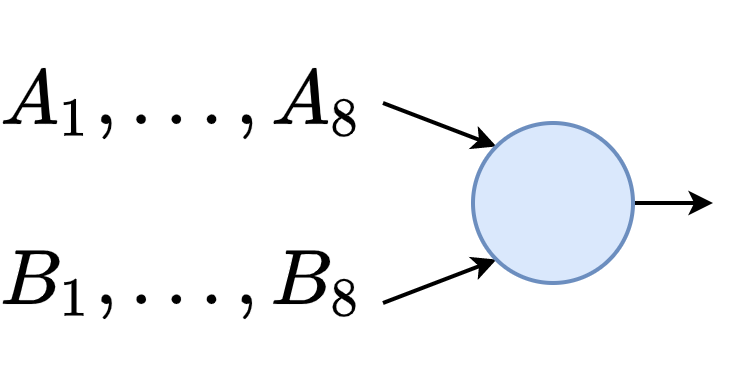}
    \caption{Task graph of the sequential reduction implementation for an example block contraction of 8 pairwise coupled blocks $A_i, B_i$. All dependency blocks must be loaded into memory to start execution.}
    \label{fig:sequential-task}
\end{figure}

Block Tensor Contraction can be seen as a kind of \textit{map-reduce} algorithm in which the contraction of pairwise-coupled blocks correspond to the \textit{mapping} and the sum of partial blocks as the \textit{reduction}. In this work, we are interested in the reduction procedure.
In general, this task is performed using a sequential implementation in which all the pairwise-coupled contractions and the reduction are done in the same task. This eases the implementation and the task scheduling, but has a high memory requirement. As the whole set of pairwise-coupled blocks are loaded into the same memory (Figure~\ref{fig:sequential-task}), the memory usage grows linearly with the dimension of the contracted indices. More specifically, the amount of memory in number of elements required by the sequential implementation is,
\begin{equation}
    \mathcal{B}_k k_b (m_b + n_b) + m_b n_b
    \label{eq:impl:mem:seq}
\end{equation}
where $m_b, n_b$ are the block dimensions of the non-contracting indices, $k_b$ is the block dimension of the contracting or common indices, and $\mathcal{B}_k$ is the number of blocks of the contracting indices which is equal to the number of pairs of blocks. The first factor corresponds to the size of the input blocks and the second factor is the size of the resulting block or \textit{accumulator}. Notice that the product $\mathcal{B}_k k_b$ is actually the dimension of the contracting indices, not the block dimension. So, if dimension of the contracting indices is big enough, input blocks may not fit in memory, and thus, turn this implementation into impossible to execute.

\begin{figure}
    \centering
    \includegraphics[width=.45\textwidth]{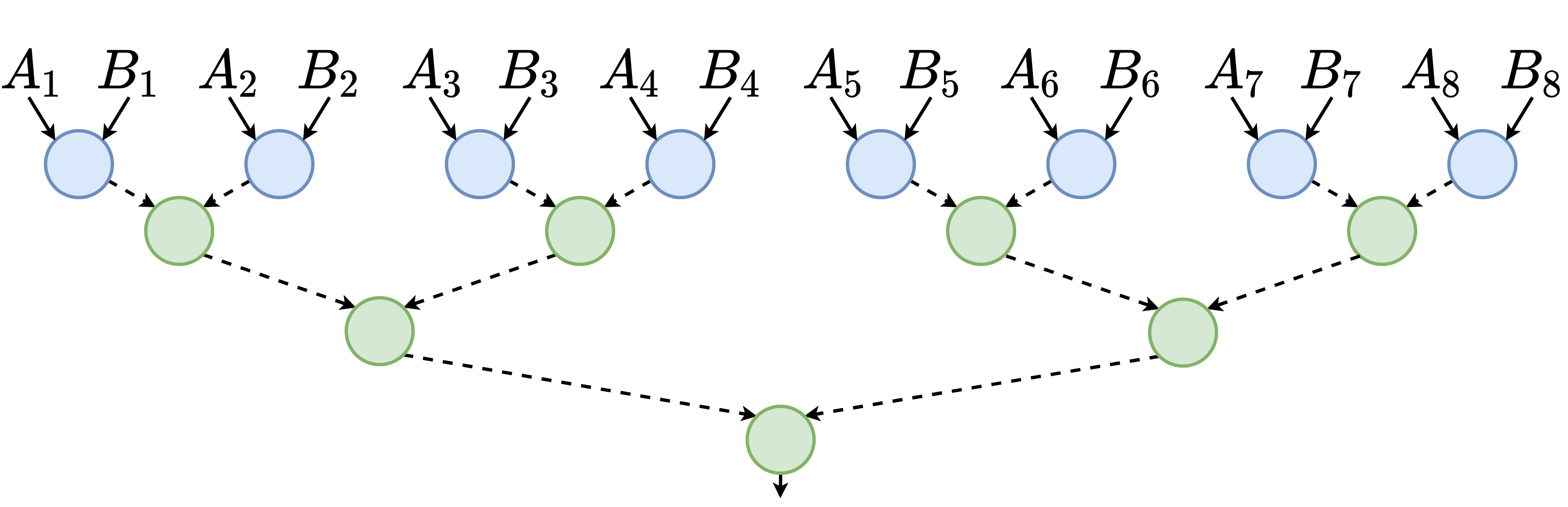}
    \caption{Task graph of the tree reduction implementation. Blue nodes perform tensor contraction and reduction, while green nodes perform just the reduction. Dashed arrows indicate the partial \textit{accumulator} dependencies.}
    \label{fig:tree-reduction}
\end{figure}

An alternative solution is to parallelize pairwise contractions and perform a tree-like reduction (Figure~\ref{fig:tree-reduction}). This method increases parallelism while the memory usage of each task shrinks to a single pairwise block contraction, where the memory usage is
\begin{equation}
    k_b (m_b + n_b) + m_b n_b
    \label{eq:impl:mem:tree-reduction}
\end{equation}

But the reduction in memory usage comes with an increased usage of the secondary storage, as replicas with partial results of the result block must coexist. Although partial blocks are discarded after reduction, in a worst-case scenario where the tree-like reduction is performed to the first-level (blue nodes in Figure~\ref{fig:tree-reduction}), the use of secondary storage is multiplied by $\frac{\mathcal{B}_k}{2}$. Furthermore, communication overhead increases significantly with an increment of serializations by a factor of $\mathcal{B}_k - 1$ and deserializations by a factor of $2 - \frac{2}{\mathcal{B}_k}$. As our goal is to squeeze all available space for quantum state representation, we opt not to parallelize this part.

We propose a variation of the sequential implementation in which the pairwise-coupled blocks are loaded as needed. We do so by sequentially concatenating tasks that perform a pairwise-coupled block contraction and the reduction of the resulting block with the \textit{accumulator} (Figure~\ref{fig:commutative-reduction}). This way the memory usage per task is as low as the parallel reduction implementation (Equation~\ref{eq:impl:mem:tree-reduction}) with no increase in secondary storage usage.
Theoretically, the amount of serializations increases by a factor of $\mathcal{B}_k$ compared to the sequential implementation due to the accumulator being constantly loaded and written. 
However, extra serializations can be avoided by just storing the accumulator in memory although we avoid this to benefit from the fault-tolerance mechanisms of COMPSs.

Even though tensor contraction is not commutative, the summation of tensors does commute. We exploit this fact by appropriately marking the \textit{accumulator} variable. The COMPSs runtime schedules the contraction of a pair of coupled blocks at the moment its dependencies are resolved, modifying the order of the reduction to minimize idle time.
Furthermore, it provides of more checkpoint opportunities making this method more resistant to faults.
As far as we are aware, the commutative reduction of the partial tensor contractions is a novel contribution. 

\subsection{Conservative Parallelization Tuning}
One way to increase parallelism is to continue slicing and shrinking the block size, thus increasing the grid, until all cores are filled with work. This comes at the cost of an exponential growth in the number of tasks and higher overhead due to data serialization and synchronization pauses. The commutative contraction procedure may palliate the synchronization delays, but the exponential number of tasks overloads the task scheduler.
Furthermore, time spent at user code decreases directly proportional to block size so after some threshold, serialization overhead surpasses execution time.

Alternatively, we may parallelize tasks internally, assigning them multiple cores.
We decide to use a conservative approach such that they use the minimum amount of cores that fit the memory.
Using Equations~\ref{eq:impl:mem:seq} and \ref{eq:impl:mem:tree-reduction}, a tuning system computes the memory requirements of each task and assigns a number of cores equal to the ratio of total memory to required memory.
This condition is chosen such that if a node is filled with tasks of similar memory usage, then the system does not run out of memory. Furthermore, the tuning system was designed to be configurable. For example, the user can set the number of blocks threshold at which the tuning system changes from sequential to commutative reduction.

\section{Evaluation} \label{section:evaluation}
For our experiments, we tested our developments in the MareNostrum 4 supercomputer: a 3456-node cluster, each node composed of:
\begin{itemize}
    \item $2\times$ Intel Xeon Platinum 8160 24C @ 2.1~GHz
    \item $12 \times 8$~GB DDR4-2667 DIMMS
    \item A 100~Gb Intel Omni-Path Full-Fat Tree Interconnection Network
    \item A 10~Gb Ethernet Interconnection Network
\end{itemize}

\begin{figure}
    \centering
    \includesvg[width=.45\textwidth]{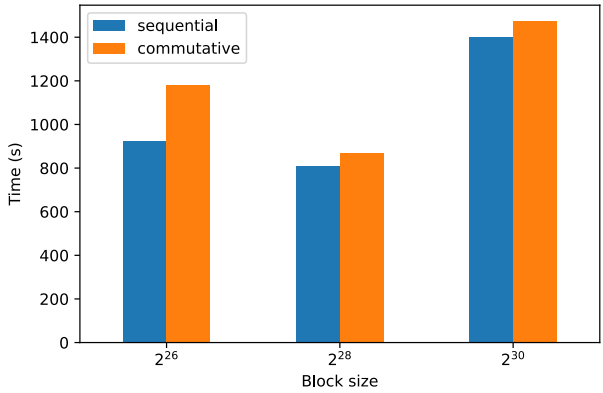}
    \caption{Execution times for the contraction of a Random Quantum Circuit with 53 qubits and depth 12, varying the maximum block size.}
    \label{fig:trace:time}
\end{figure}

Verification was performed by simulating a single exact amplitude of a Random Quantum Circuit \cite{chen_classical_2018} composed of 53 qubits and circuit depth 12, as designed for Google's Sycamore simulation \cite{arute_quantum_2019}.
We used \texttt{quimb}~\cite{gray_quimb_2018} for tensor network simplification and \texttt{cotengra}~\cite{gray_hyper-optimized_2021} for contraction path optimization.
For tensor slicing, we used \texttt{SliceFinder} algorithm included in \texttt{cotengra}.
Due to limitations on the maximum number of dimensions of arrays, we forked and adapted the \texttt{numpy} library to surpass the limit of 32 dimensions.
Notice how maximum block size tuning is critical for performance as shown in Figure~\ref{fig:trace:time}.

\begin{figure}
    \centering
    \includegraphics[width=.45\textwidth]{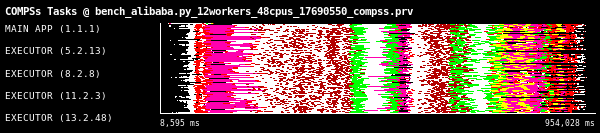}
    \caption{Task Timeline of the Random Quantum Circuit simulation \cite{chen_classical_2018}. All tasks are shown here, whether they are blocked or non-blocked. Colors are assigned by task routine and number of cores assigned.}
    \label{fig:trace:tasks-timeline}
\end{figure}

\begin{figure}
    \centering
    \includegraphics[width=.43\textwidth]{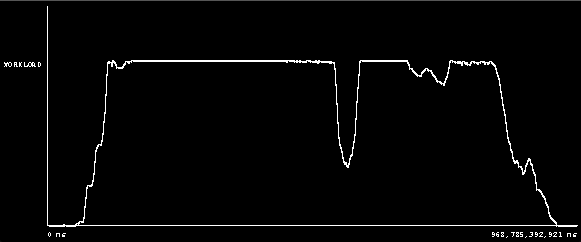}
    \caption{CPU usage along time of  the Random Quantum Circuit simulation \cite{chen_classical_2018}. The line shows the instantaneous cumulative number of CPUs used by the application }
    \label{fig:trace:cpus}
\end{figure}

Thanks to PyCOMPSs's profiling capabilities, we are able to introspect into the execution. A task timeline of the execution is shown in Figure~\ref{fig:trace:tasks-timeline}, and instantaneous use of CPUs is shown in Figure~\ref{fig:trace:cpus}. The application is able to use most of the CPUs available for tasks all the time. There is only a large reduction of the resource usage during a short time after a first reduction phase.

\begin{figure}
    \centering
    \begin{subfigure}{.45\textwidth}
        \centering
        \includesvg[width=\textwidth]{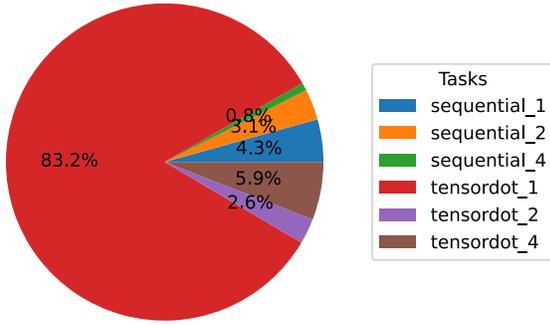}
        \caption{Sequential Reduction}
    \end{subfigure}
    \begin{subfigure}{.45\textwidth}
        \centering
        \includesvg[width=\textwidth]{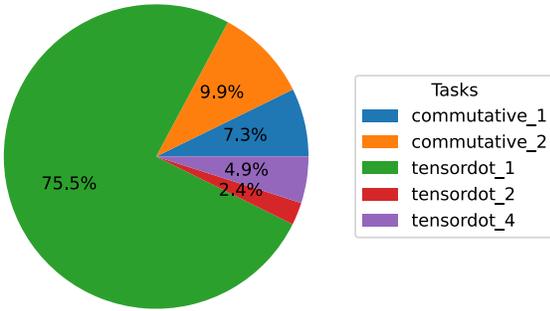}
        \caption{Commutative Reduction}
    \end{subfigure}
    \caption{Time percentage spent at each of the task routines for the simulation of the Random Quantum Circuit using (a) Sequential Reduction and (b) Commutative Reduction.}
    \label{fig:trace:events-pie}
\end{figure}

\begin{figure}
    \centering
    \includesvg[width=.45\textwidth]{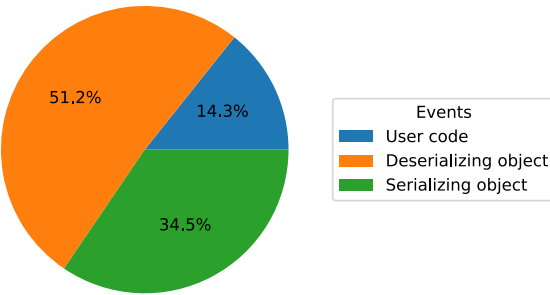}
    \caption{Time percentage spent at deserialization, serialization and user code (tensor contraction) for the \texttt{tensordot\_1} task on the execution shown in Figure~\ref{fig:trace:tasks-timeline}.}
    \label{fig:trace:events:tensordot-1}
\end{figure}

By analyzing the time spent at each task instance (Figure~\ref{fig:trace:events-pie}), we can check that the bottleneck of the execution for both of the experiments is the \texttt{tensordot} task (non-blocked tensor contraction) running on just 1 core. PyCOMPSs exports internal events and if we focus on the \texttt{tensordot\_1} task (Figure~\ref{fig:trace:events:tensordot-1}) it shows to be bounded by deserialization, taking half of the time. Notice also in Figure~\ref{fig:trace:events-pie} that \texttt{commutative} tasks are taking a higher percentage of the time that \texttt{sequential} tasks, suggesting that for this simulation the overhead of the Commutative Reduction exceeds its benefits.


\subsection{Weak Scalability Analysis}
As the size of a circuit grows, the contraction of the tensor network employs more space and more nodes are needed to solve the problem.
We perform a weak scalability analysis by measuring the execution time efficiency of the contraction of 2 matrices $U_{mk}, V_{nk}$. Their block sizes are fixed to $(m_b, k_b), (n_b, k_b)$ where $m_b = n_b = 2^{14}, k_b = 2^{13}$ single-precision complex numbers. In order to scale the workload, we increase the number of blocks $B_m, B_n$ of the non-contracting indices following the entries of Table~\ref{tab:weak:num-blocks}. Likewise, we assess the behavior of the Conservative Parallelization Tuning by varying the number of blocks of the contracted index $\mathcal{B}_k$.
We performed trials for $\mathcal{B}_k$ equal to 2, 4, 8 and 16 blocks.

\begin{table}
    \centering
    \begin{tabular}{|c|cccccc|}
        \hline
        Nodes & 1            & 2             & 4              & 8              & 16             & 32             \\ \hline
        Par.  & $8 \times 6$ & $8 \times 12$ & $16 \times 12$ & $16 \times 24$ & $32 \times 24$ & $32 \times 48$ \\ \hline
    \end{tabular}
    \caption{Parallelism of the tensor contraction in number of blocks for the non-contracting indices $B_m,B_n$. }
    \label{tab:weak:num-blocks}
\end{table}

\begin{figure}
    \centering
    \begin{subfigure}{.45\textwidth}
        \centering
        \includesvg[width=\textwidth]{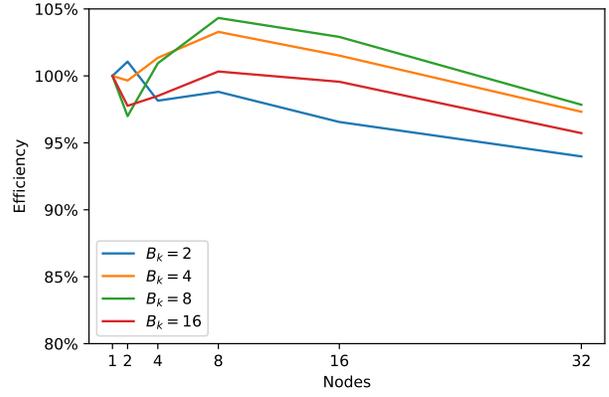}
        \caption{Sequential Reduction}
    \end{subfigure}
    \begin{subfigure}{.45\textwidth}
        \centering
        \includesvg[width=\textwidth]{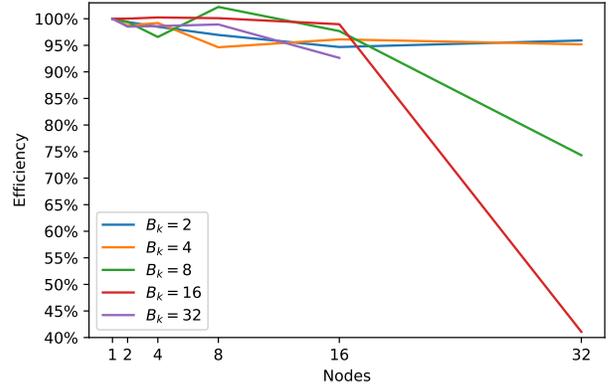}
        \caption{Commutative Reduction}
    \end{subfigure}
    \caption{Execution efficiency with scaling number of nodes for number of blocks in contracting index $\mathcal{B}_k = 2, 4, 8, 16$ and $32$.}
    \label{fig:weak:nodes}
\end{figure}

In the case of the sequential reduction, the required memory is directly proportional to $\mathcal{B}_k$ so the tuning should take effect by requesting more cores per task in the same proportion. We expect that as the granularity of the tuned tasks completely fit the available cores, then we see a performance similar to the commutative reduction case.
In Figure~\ref{fig:weak:nodes}, the system holds a $90-95\%$ efficiency on both implementations for $\mathcal{B}_k \leq 4$ and 32 nodes. Nevertheless, the commutative reduction shows a drastic efficiency drop for $\mathcal{B}_k = 8$ ($75\%$) and $\mathcal{B}_k = 16$ ($40\%$). We have analysed tracefiles of the executions and find out a scheduling issue in the COMPSs runtime that appears in the 32 nodes case with large values of $\mathcal{B}_k$. We are working on improvements on the scheduler to avoid this efficiency drop.

\begin{figure}
    \centering
    \begin{subfigure}{.45\textwidth}
        \centering
        \includesvg[width=\textwidth]{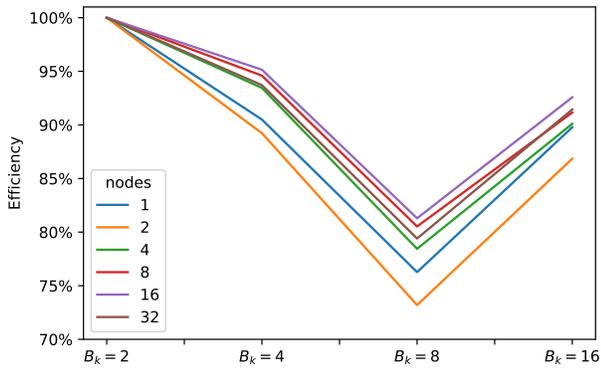}
        \caption{Sequential Reduction}
        \label{fig:weak:bk:seq}
    \end{subfigure}
    \begin{subfigure}{.45\textwidth}
        \centering
        \includesvg[width=\textwidth]{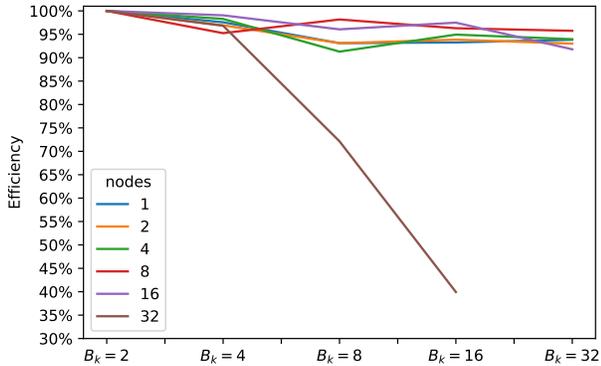}
        \caption{Commutative Reduction}
        \label{fig:weak:bk:comm}
    \end{subfigure}
    \caption{Execution efficiency with scaling number of blocks in the contracting index.}
\end{figure}

By measuring the efficiency as a function of the number of blocks in the contracting index $\mathcal{B}_k$, a new point of view is provided.
In Figure~\ref{fig:weak:bk:seq}, the sequential implementation shows a 20-30\% efficiency loss at $\mathcal{B}_k=8$.
At $\mathcal{B}_k=32$, the sequential reduction runs out of memory and fails while the commutative reduction continues working.
In Figure~\ref{fig:weak:bk:comm}, the commutative reduction shows a near perfect scalability for all number of blocks on the contracting index $\mathcal{B}_k$ and number of nodes, except for the case of using 32 nodes. Notice that unlike the sequential implementation, the commutative reduction is able to scale up to $\mathcal{B}_k = 32$.


% ...

\section{Related Work} \label{section:related_work}
Available distributed tensor-network simulators are ExaTENSOR~\cite{liakh_dmytro_exatensor_2019} and Jet~\cite{vincent_jet_2021}. An interesting library for performing distributed tensor contractions is Cyclops~\cite{solomonik_cyclops_nodate}. \rosnet\ is a pure Python library focused on distribution, while local execution is delegated to specialized libraries developed by the community thanks to the \texttt{numpy}'s dispatching mechanism.

Task-based programming has been proposed as an alternative to parallel programming~\cite{thoman2018taxonomy}. It offers multiple features such as asynchronous execution, load balancing and improving programmability. Due to the nature of classic quantum computing simulation, which is a hungry application both in terms of compute and storage, we are interested in a distributed instance of such. Also, we are interested on environments that support programming in Python. Between the alternative task-based environments available in Python we find Dask~\cite{rocklin_dask_2015}, Ray~\cite{moritz_ray_nodate}, Parsl~\cite{babuji2019parsl} and Pygion~\cite{slaughter2019pygion}. The main features that differentiate PyCOMPSs are:
\begin{itemize}
    \item Data is located out-of-core.
          Our benefits here are two-fold: (1) Fault-tolerance is straightforward and ensured for each task and (2) with a higher secondary storage, the size of the quantum states we can represent grows.
          Main memory is only used for task execution and caching data.
    \item Detailed tracing and profiling support with Extrae and Paraver \cite{pillet1995paraver}.
          The runtime can register hardware counters, (de)serialization bandwidth, and a large plethora of runtime and user events that can then be visualized in a timeline view.
    \item Dynamic Heterogeneous Computing. The user may provide several implementations for the same routine and COMPSs will select the implementation on runtime based on the available resources. This way we may use hardware accelerators without ahead-of-time planning.
\end{itemize}

\section{Conclusions} \label{section:conclusions}
In this work, we present the \rosnet\ library for distributed tensor algebra. Using the COMPSs runtime, our library features out-of-core data location, tracing and profiling support, and potential for heterogeneous computing. Thanks to \textit{numpy}'s dispatching mechanism, we can leverage over \textit{numpy} derivatives. 

We developed a new block contraction method that uses substantially less memory per task and thus can scale further than traditional block contraction implementations. Furthermore, the library lets the user to choose the implementation by configuring the tuning system.
Our library is a novel contribution bringing flexible tools to perform large-scale simulations of Quantum systems to a community of Quantum developers.

Future work will focus on further expanding the available routines, such as tensor decompositions and optimized contraction methods, and runtime development focused on improving the communication backend and the scheduling system.

\section*{Acknowledgment}
We acknowledge support from project QuantumCAT (ref. 001- P-001644), co-funded by the Generalitat de Catalunya and the European Union Regional Development Fund within the ERDF Operational Program of Catalunya, and European Union’s Horizon 2020 research and innovation programme under grant agreement No 951911 (AI4Media). This work has also been partially supported by the Spanish Government (PID2019-107255GB) and by Generalitat de Catalunya (contract 2014-SGR-1051). This work is co-funded by the European Regional Development Fund under the framework of the ERFD Operative Programme for Catalunya 2014-2020, with 1.527.637,88€. Anna Queralt is a Serra H\'unter Fellow.

\bibliographystyle{ieeetr}
\bibliography{quantic, bibliography}

\end{document}